\documentclass[11pt,en]{elegantpaper}

\newcommand{\jelcodes}[1]{\vskip 0.5ex\par
  \noindent\normalfont\bfseries JEL~codes: #1}

\usepackage{amsmath,amssymb}
\usepackage{booktabs}
\usepackage{graphicx}
\usepackage{hyperref}
\usepackage{csquotes}

\usepackage{array}
\usepackage{algorithm}
\usepackage{algpseudocode}
\usepackage{subcaption}
\usepackage{threeparttable}

\title{The Aligned Economic Index \& the State-Switching Model}
\author{Ilias Aarab}
\institute{European Systemic Risk Board, European Central Bank\footnote{\emph{Disclaimer: This paper should not be reported as representing the views of the European Systemic Risk Board (ESRB) or the European Central Bank (ECB). The views expressed are those of the authors and do not necessarily reflect those of the ESRB or ECB.}}\\University of Antwerp}
\date{June 2019}

\begin{document}

\maketitle

\begin{abstract}
A growing empirical literature suggests that equity-premium predictability is state dependent, with much of the forecasting power concentrated around recessionary periods \parencite{Henkel2011,DanglHalling2012,Devpura2018}. I study U.S. stock return predictability across economic regimes and document strong evidence of time-varying expected returns across both expansionary and contractionary states. I contribute in two ways. First, I introduce a \textbf{state-switching predictive regression} in which the market state is defined in real time using the slope of the yield curve. Relative to the standard one-state predictive regression, the state-switching specification increases both in-sample and out-of-sample performance for the set of popular predictors considered by \textcite{WelchGoyal2008}, improving the out-of-sample performance of most predictors in economically meaningful ways. Second, I propose a new aggregate predictor, the \textbf{Aligned Economic Index}, constructed via partial least squares (PLS). Under the state-switching model, the Aligned Economic Index exhibits statistically and economically significant predictive power in sample and out of sample, and it outperforms widely used benchmark predictors and alternative predictor-combination methods.
\par
\jelcodes{G12, G17, E44, C22, C53}
\end{abstract}

\section{Introduction}

A large body of work studies whether stock returns are predictable using valuation ratios and macroeconomic indicators (e.g., \parencite{FamaFrench1988,CampbellShiller1988,KandelStambaugh1996,Guo2006,Lewellen2004,PolkThompsonVuolteenaho2006}). \textcite{WelchGoyal2008} show, however, that many prominent predictive regressions fail to deliver reliable out-of-sample gains relative to simple benchmarks. Two recurrent explanations are model uncertainty, the correct specification is unknown ex ante. and parameter instability, which makes estimates sensitive to the sample period.

Over the past decades, several approaches have been proposed to mitigate these issues, including economically motivated restrictions \parencite{PanPettenuzzoWang2020}, combinations of multiple predictors \parencite{Huang2015}, regime-shift specifications \parencite{HammerschmidLohre2018}, and new predictors \parencite{Jiang2019}. While these methods often improve forecasting performance, their gains are frequently concentrated in recessionary episodes, with weaker results in expansions.

A key mechanism behind this pattern is that investors may rely on different information sets and forecasting rules across states. \textcite{CujeanHasler2017} provide an equilibrium framework in which agents switch attention and forecasting models over time, and \textcite{Devpura2018} document predictor-dependent time variation in return predictability. Motivated by these findings, I use a simple state-dependent predictive regression in which slope coefficients are allowed to change across market states. The resulting state-switching model delivers stronger and more stable predictive performance across both expansionary and recessionary periods than the conventional one-state specification.

\section{Data}

The aggregate stock market return is the excess return on the S\&P~500 index: the continuously compounded log return (including dividends) minus the risk-free rate, proxied by the three-month Treasury bill. Working with excess returns focuses attention on the predictability of the (real) equity risk premium rather than on variation driven mechanically by the level of short-term interest rates.

I use updated data from \textcite{WelchGoyal2008} covering January 1950 to December 2017. The predictor set includes the 14 popular variables in \textcite{WelchGoyal2008}, supplemented with the corporate bond premium and the lagged equity premium, for a total of 16 series.\footnote{The predictors are: dividend--price ratio (log), dividend yield (log), earnings--price ratio (log), dividend--payout ratio (log), equity premium volatility, book-to-market ratio, net equity expansion, Treasury bill rate, long-term yield, long-term return, term spread, default yield spread, default return spread, inflation, lagged equity premium, and the corporate bond premium. The data can be retrieved from Amit Goyal’s web page (\texttt{www.hec.unil.ch/agoyal/}); variable definitions are provided in \textcite{WelchGoyal2008}.}
Following common practice, I focus on the post-war period. From an investor-relevance perspective, this emphasizes modern market dynamics. From a statistical perspective, \textcite{Lewellen2004} argues that pre-1945 data have markedly different properties (e.g., unusually high volatility and persistence in some predictors), which can distort inference. I also exclude 1945--1949 because dividend policies during and immediately after World War II were unusually volatile, potentially affecting the behavior of dividend-based predictors \parencite{Frankfurter1997}. Finally, based on preliminary analysis on the first years of data, the baseline sample used throughout is January 1960 to December 2017 (696 monthly observations).

All data are monthly and the baseline forecasting horizon is one month ahead. This horizon is well suited for studying state dependence: business-cycle states typically last several months, so longer-horizon regressions may mechanically average over multiple states and blur regime effects. Moreover, short-horizon predictability can translate into longer-horizon predictability under standard present-value logic \parencite{Cochrane2011,Huang2015,RapachRinggenbergZhou2016}.

To define market states in real time, I construct an ex-ante indicator based on the slope of the yield curve (the term spread), computed as the 10-year Treasury yield minus the 3-month Treasury bill rate. In contrast to NBER recession dates, which are assigned ex post,\footnote{NBER recession dates can be retrieved from the Federal Reserve Bank of St.\ Louis (FRED). Because these dates are determined ex post, they are not directly usable for real-time forecasting exercises.} the yield-curve slope is observable at the time forecasts are formed and is widely viewed as a leading indicator.

\section{Methodology}

\subsection{Return predictability and switching between states}

Under the expectations hypothesis, the term spread reflects the difference between the current short rate and expected future short rates over a longer horizon. When the spread narrows, investors may be more uncertain about future economic conditions and require higher compensation for risk. The yield curve therefore summarizes aspects of macroeconomic sentiment and risk compensation \parencite{Wright2006}.

The slope of the Treasury yield curve is often treated as an early-warning indicator for downturns: inversions have historically preceded several U.S.\ recessions. Motivated by this, I define the market state at month $t$ as a \emph{down state} if the yield curve was flat or inverted at any point in the preceding $\tau$ months:
\begin{equation}\label{eq:state_indicator}
S_t =
\begin{cases}
1, & \text{if } \text{tms}_{t-i} \le 0 \text{ for some } i \in \{1,\dots,\tau\},\\
0, & \text{otherwise},
\end{cases}
\end{equation}
where $S_t$ is the state indicator and $\text{tms}_{t-i}$ is the term spread (10-year minus 3-month), as in, e.g., \parencite{FamaFrench1989,WelchGoyal2008}. The parameter $\tau$ controls how long a down state persists after a (near-)inversion. To reduce the risk of in-sample overfitting and data-snooping concerns emphasized by \textcite{WelchGoyal2008}, I fix $\tau$ rather than optimizing it. In the baseline results, I set $\tau=9$ months.\footnote{Using $\tau$ between 3 and 12 months leaves the main findings unchanged.} This choice is short enough to avoid mechanically locking the economy into long down states, while remaining consistent with the average duration of recessionary episodes observed in the data.

\subsection{The state-switching model}

I extend the standard predictive regression by allowing slope coefficients to differ across states, as in \parencite{Boyd2005,Sander2018}:
\begin{equation}\label{eq:switching_model}
R_{t+1} = (\beta_0 + \delta_0 S_t) + \beta_1 S_t x_t + \gamma_1 (1 - S_t) x_t + \varepsilon_{t+1},
\end{equation}
where $R_{t+1}$ is the excess return on the S\&P~500 at month $t+1$, $x_t$ is a lagged predictor observed at month $t$, and $\varepsilon_{t+1}$ is the disturbance term. When $\beta_1=\gamma_1$, the model collapses to the conventional one-state predictive regression.

Out-of-sample forecasts are formed recursively:
\begin{equation}\label{eq:recursive_forecast}
r_{t+1|t} = (b_0 + d_0 S_t) + b_1 S_t x_t + g_1 (1 - S_t) x_t,
\end{equation}
where $(b_0,d_0,b_1,g_1)$ are OLS estimates of $(\beta_0,\delta_0,\beta_1,\gamma_1)$ computed using only information available up to time $t$. Specifically, for each forecast origin $t$, I regress $\{R_{s+1}\}_{s=M}^{t-1}$ on $\{1, S_s, S_s x_s, (1-S_s)x_s\}_{s=M}^{t-1}$, where $M$ is the start of the out-of-sample period. This recursive procedure avoids look-ahead bias by construction.

Combining complementary information from multiple predictors can yield strong forecasting performance across different market states. However, as shown by \textcite{WelchGoyal2008}, estimating a multivariate ``kitchen-sink'' regression typically leads to poor out-of-sample performance due to parameter instability and overfitting. To address this, I aggregate information from the 16 predictors in a more parsimonious way by constructing an aligned economic index.

The economic intuition is that investors may shift attention across fundamentals and macro indicators depending on market conditions. Consistent with this view, changes in investor attention and information processing have been linked to return dynamics \parencite{BarberOdeanZhu2009,DaEngelbergGao2011}.

\subsection{The Aligned Economic Index}

A standard approach to summarize many predictors is principal component analysis (PCA), which reduces dimensionality by extracting common factors that explain maximal predictor variance \parencite{StockWatson2002,Jolliffe2002}. When predictors are noisy proxies and idiosyncratic components are correlated across series, however, PCA factors may be contaminated by common noise and become less useful for forecasting \parencite{BoivinNg2006}. Partial least squares (PLS) provides an alternative: it constructs components that explicitly maximize covariance with the forecast target (here, future excess returns), rather than explaining total predictor variance \parencite{GeladiKowalski1986}. This targeted dimension reduction is well suited to forecasting settings in which much predictor variation may be irrelevant for prediction \parencite{KellyPruitt2015}. Appendix~\ref{app:aei} describes the construction of the Aligned Economic Index in detail.

\begin{figure}[httb]
  \centering
  \includegraphics[width=0.8\linewidth]{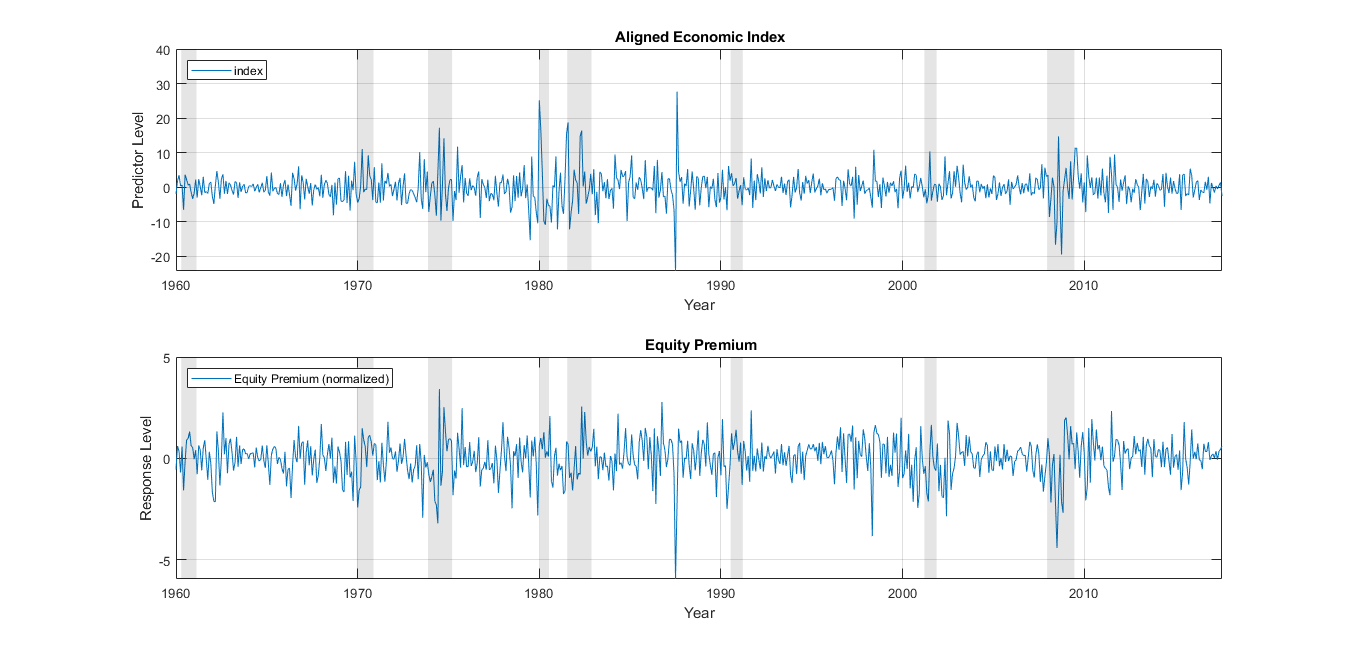}
  \caption{The Aligned Economic Index and the equity premium. The upper panel depicts the Aligned Economic Index $E^{PLS}$; the lower panel shows the normalized equity premium. The sample period is January 1960 to December 2017. Vertical bars denote NBER-dated recessions.}
  \label{fig:aei_equity_premium}
\end{figure}

Figure~\ref{fig:aei_equity_premium} plots the Aligned Economic Index (upper panel) and the normalized equity premium (lower panel) over January 1960 to December 2017. The gray bars indicate NBER recessions. If the index contains forecasting information about the risk premium, the two series should comove at business-cycle frequencies.

The figure suggests substantial comovement. Both series drop sharply early in the sample, rise during subsequent recoveries, and display heightened volatility around major market stress episodes (e.g., the late-1987 crash and the 2007--2009 financial crisis). The index is also noticeably less persistent than several classical valuation ratios, suggesting that small-sample bias concerns emphasized by \textcite{Stambaugh1999} are less acute for this predictor.

\section{Empirical Results}

Table~\ref{tab:combination} reports predictive regression results using the Aligned Economic Index. Panel A estimates the conventional one-state model using $E^{PLS}$. Panel B estimates the state-switching model in Equation~\eqref{eq:switching_model}. For comparison, I also report results for two alternative predictor-combination methods: (i) a PCA-based factor $E^{PCA}$,\footnote{I retain only the first principal component. In unreported results, including additional factors consistently reduces out-of-sample performance in both the one-state and state-switching specifications.} and (ii) the forecast-combination (FC) approach of \textcite{Rapach2010}, denoted $E^{FC}$.

If return predictability is time varying and predictor dependent \parencite{Devpura2018}, one would expect that allowing coefficients to switch across states should improve performance relative to the one-state specification, and that the magnitude of the gains may differ across combination methods.

\begin{table}[h]
\centering
\caption{Forecasting the market excess return with predictor-combination methods}
\label{tab:combination}

\begin{tabular}{lcccc}
\toprule
\multicolumn{5}{c}{Panel A: One-state model} \\
\midrule
Predictor & $\beta_1$ & t-stat & $\bar R^2$ & $R^2_{\text{oos}}$ \\
\midrule
$E^{PLS}$ & 0.05*** & 4.38 & 4.65  & 2.60*** \\
$E^{PCA}$ & 0.20**  & 1.96 & 0.61  & 0.31    \\
$E^{FC}$  & 0.13    & 0.84 & -0.08 & 1.18*   \\
\bottomrule
\end{tabular}

\vspace{0.9em}

\begin{tabular}{lcccccccc}
\toprule
\multicolumn{9}{c}{Panel B: State-switching model} \\
\midrule
Predictor
& $\delta_0$ & t-stat
& $\beta_1$  & t-stat
& $\gamma_1$ & t-stat
& $\bar R^2$ & $R^2_{\text{oos}}$ \\
\midrule
$E^{PLS}$ & -0.27*** & -2.51 & 0.03*** & 3.46 & 0.06*** & 3.68 & 5.90 & 4.12*** \\
$E^{PCA}$ & -0.01*** & -2.54 & 0.17    & 1.27 & 0.25**  & 1.97 & 1.65 & 1.58*** \\
$E^{FC}$  & -1.22*** & -2.43 & 0.24    & 0.85 & 0.15    & 0.81 & 1.02 & 2.64*** \\
\bottomrule
\end{tabular}

\vspace{0.5em}
\small This table reports one-month-ahead forecasts of the market excess return using $E^{PLS}$, $E^{PCA}$, and $E^{FC}$. Panel A estimates the standard one-state predictive regression. Panel B estimates the state-switching model in Equation~\eqref{eq:switching_model}. In Panel B, $S_t=1$ when the yield curve was flat or inverted at least once in the preceding nine months, and $S_t=0$ otherwise. $\bar R^2$ is the in-sample adjusted $R^2$ over January 1960 to December 2017. $R^2_{\text{oos}}$ is the out-of-sample $R^2$ of \textcite{CampbellThompson2008}, using the first 20 years as the initial estimation window and January 1980 to December 2017 as the evaluation period. t-statistics are Newey--West. Statistical significance for $R^2_{\text{oos}}$ is based on the MSFE-adjusted statistic of \textcite{ClarkWest2007}. ***, ** and * indicate significance at the 1\%, 5\%, and 10\% levels, respectively.
\end{table}

Panel A shows that both $E^{PLS}$ and $E^{PCA}$ have significant in-sample predictive power. The adjusted $\bar R^2$ of $E^{PLS}$ (4.65\%) is notably larger than that of $E^{PCA}$ (0.61\%), indicating that $E^{PLS}$ explains a larger share of one-month-ahead variation in the equity premium. The forecast-combination predictor $E^{FC}$ has weak in-sample performance in this specification. Out of sample, $E^{PCA}$ yields a positive but small $R^2_{\text{oos}}$ (0.31\%) that is not statistically significant and falls below the 0.50\% economic threshold discussed by \textcite{CampbellThompson2008}. In contrast, $E^{FC}$ achieves a higher $R^2_{\text{oos}}$ (1.18\%), consistent with the forecasting gains reported by \textcite{Rapach2010}, while $E^{PLS}$ performs best with $R^2_{\text{oos}}=2.60\%$.

Panel B turns to the state-switching model. Two patterns stand out. First, the state intercept shift $\delta_0$ is statistically significant (Newey--West t-statistics around 2.5) across all three predictors, supporting the relevance of yield-curve-based regimes. Second, both $\bar R^2$ and $R^2_{\text{oos}}$ increase relative to the one-state model for all three predictor-combination methods. All three predictors now produce statistically significant $R^2_{\text{oos}}$ values that comfortably exceed the \textcite{CampbellThompson2008} 0.50\% threshold. Among them, $E^{PLS}$ remains the strongest performer, with $\bar R^2=5.90\%$ and $R^2_{\text{oos}}=4.12\%$.

\begin{table}[h]
\centering
\caption{Forecasting the market excess return across different states}
\label{tab:states}
\begin{tabular}{lcccccccc}
\toprule
 & \multicolumn{4}{c}{Expansions/Recessions} & \multicolumn{4}{c}{Up/Down states} \\
\cmidrule(lr){2-5}\cmidrule(lr){6-9}
Predictor & $\bar R^2_{\text{exp}}$ & $\bar R^2_{\text{rec}}$ & $R^2_{\text{oos,exp}}$ & $R^2_{\text{oos,rec}}$ &
$\bar R^2_{\text{up}}$ & $\bar R^2_{\text{down}}$ & $R^2_{\text{oos,up}}$ & $R^2_{\text{oos,down}}$ \\
\midrule
\multicolumn{9}{c}{Panel A: One-state model} \\
\midrule
$E^{PLS}$ & 0.57 & 14.49 & 1.45** & 5.99*** & 5.51 & 2.83 & 4.31*** & -7.86 \\
$E^{PCA}$ & -0.70 & 4.26 & -1.36 & 5.24*** & 0.57 & 1.27 & 0.05 & 1.88 \\
$E^{FC}$  & 0.03 & 0.14 & 0.77 & 2.38*** & 0.06 & 0.07 & 1.24** & 0.80 \\
\midrule
\multicolumn{9}{c}{Panel B: State-switching model} \\
\midrule
$E^{PLS}$ & 1.01 & 19.13 & 1.71**  & 11.21*** & 6.06 & 6.98 & 4.51*** & 1.68** \\
$E^{PCA}$ & -0.60 & 8.50 & 0.00    & 6.23***  & 0.92 & 5.18 & 0.67**  & 7.12*** \\
$E^{FC}$  & 0.02 & 4.92 & 2.19***  & 3.95***  & 0.40 & 4.29 & 2.03*** & 6.36*** \\
\bottomrule
\end{tabular}

\vspace{0.5em}
\small This table reports state-by-state forecasting performance for the one-state predictive regression (Panel A) and the state-switching model in Equation~\eqref{eq:switching_model} (Panel B). $E^{PLS}$, $E^{PCA}$, and $E^{FC}$ denote the PLS-based Aligned Economic Index, the PCA-based factor, and the forecast-combination predictor, respectively. $\bar R^2_{\text{exp}}$ and $\bar R^2_{\text{rec}}$ are in-sample adjusted $R^2$ computed within NBER expansions and recessions. $\bar R^2_{\text{up}}$ and $\bar R^2_{\text{down}}$ are in-sample adjusted $R^2$ computed within yield-curve up and down states as defined by $S_t$. $R^2_{\text{oos,}\cdot}$ denotes the \textcite{CampbellThompson2008} out-of-sample $R^2$ computed within each state, using 1960--1979 as the initial estimation window and 1980--2017 as the evaluation period. Statistical significance for $R^2_{\text{oos}}$ is based on the \textcite{ClarkWest2007} MSFE-adjusted statistic. ***, ** and * indicate significance at the 1\%, 5\%, and 10\% levels.
\end{table}

Although $E^{PLS}$ is a strong predictor of excess monthly returns when looking at the entire sample period, it is interesting to further evaluate its performance during certain market states. Table~\ref{tab:states} reports the results across different market states. Panel A shows the forecasting performance using the conventional one-state univariate model, while panel B reports findings using the state switching model. The left part of Panel A shows the in-sample and out-of-sample performance of the one-state model across expansions and recessions. Both $E^{PLS}$ and $E^{PCA}$ can forecast, in-sample, the equity premium during recessions, however only $E^{PLS}$ is able to generate a positive adjusted $\bar R^2$ during expansions. Looking at the out-of-sample performance, both $E^{PCA}$ and $E^{FC}$ produce significant positive $R^2_{\text{oos}}$ values during recessions, but fail to do so during expansions, with $E^{PCA}$ even failing to outperform the benchmark model of the historical mean. These findings are in line with the conclusions of \cite{Henkel2011} and \cite{DanglHalling2012}: return predictability only exists during recessionary periods. In contrast, the newly formed aligned economic index can significantly predict the equity premium during recessions and expansions. $E^{PLS}$ has an $R^2_{\text{oos}}$ of 5.99\% during recessions and an $R^2_{\text{oos,exp}}$ of 1.45\% during expansions, both significant at the 5\% level or stronger. This confirms that using combined information of the predictors in an efficient way leads to improvement across different states. The right part of Panel A shows the same statistics across up and down states as defined by the state indicator based on the inversion of the yield curve (as defined in Equation \eqref{eq:state_indicator}). All three models produce positive in-sample $\bar R^2$ across both states, with $E^{PLS}$ outperforming the models substantially across both states. The out-of-sample results are less clear. First, $E^{PCA}$ fails to significantly outperform the benchmark model during an upstate and during a downstate. Thus, $E^{PCA}$ is only able to capture predictability during parts of the recessionary periods that do not overlap with the downstate periods. Secondly, both $E^{PLS}$ and $E^{FC}$ outperform, significantly on the 5\% level or stronger, the naïve benchmark model during an upstate but fail to do so during a downstate with $E^{PLS}$ even generating a negative $R^2_{\text{oos,down}}$ of -7.86\%. A downstate is initiated whenever the yield curve inverts, which generally happens right before the start of a recession. Thus, with the one-state model $E^{PLS}$ fails to predict the equity premium right before the start of a recession. These findings are in line with \cite{Sander2018} who shows that estimating the start of a recession inaccurately can lead to large economic consequences. In short, even though $E^{PLS}$ outperforms the other two combination methods across different states, it still fails to outperform the naïve benchmark model around the start of a recession in the one-state model.

Moving on to the state switching model in panel B of Table~\ref{tab:states}, we see substantial performance gains across different states. First the in-sample performance of all three combination methods are further enhanced within the state switching model, however in line with panel A only $E^{PLS}$ is able to generate performance gains exceeding 1\% across all states. More noteworthy, the out-of-sample performance is substantially affected by the state switching model. $E^{PCA}$ is able to significantly forecast the equity premium during recessions, up states and down states. During expansions however both $E^{PCA}$ and the benchmark model predict, on average, the equity premium with the same capacity producing an $R^2_{\text{oos}}$ of zero. In contrast, both $E^{PLS}$ and $E^{FC}$ can significantly, on the 5\% level or stronger, predict the equity premium during expansions, recessions, up states and down states. $E^{PLS}$ generates $R^2_{\text{oos}}$ values of respectively 1.71\% and 1.68\% during expansions and down states, while producing $R^2_{\text{oos}}$ values of respectively 11.21\% and 4.51\% during recessions and up states. All of the values thus exceed the economic threshold of 0.50\% of \cite{CampbellThompson2008} meaning that an investor using the state switching model with $E^{PLS}$ would make consistent substantial gains across the different market states. The main difference with the one-state model is the fact that $E^{PLS}$ is now able to predict the equity premium during the recessionary turning points as well, with the inversion of the yield curve (i.e. a down state) helping to time the right moment to change slope coefficients. Interestingly, the forecast combination method $E^{FC}$ exhibits the same overall behavior as $E^{PLS}$. $E^{FC}$ generates $R^2_{\text{oos}}$ values of respectively 2.19\% and 6.36\% during expansions and down states, while producing $R^2_{\text{oos}}$ values of respectively 3.95\% and 2.03\% during recessions and up states. Thus both $E^{PLS}$ and $E^{FC}$ are able to consistently outperform the naïve benchmark model across all market states, with $E^{PLS}$ yielding the overall greatest performance. In the next section I further analyze the relationship between $E^{PLS}$ and $E^{FC}$.

\subsection{$E^{PLS}$ versus $E^{FC}$}

Over the out-of-sample period (January 1980 to December 2017), the forecasts implied by $E^{PLS}$ and $E^{FC}$ are strongly correlated (about 65\%). The correlation rises further within down states, suggesting that both methods capture a similar component of risk-premium variation when turning-point risk is elevated.

\begin{figure}[h]
  \centering
  \includegraphics[width=0.8\linewidth]{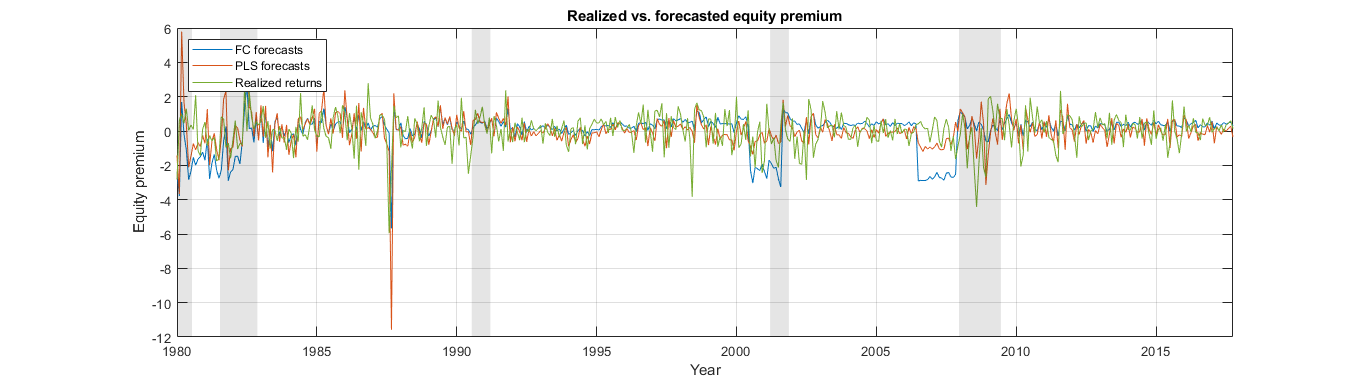}
  \caption{Excess market return forecasts of $E^{PLS}$ and $E^{FC}$, January 1980 to December 2017. The orange line shows the out-of-sample forecast based on $E^{PLS}$, the blue line the forecast based on $E^{FC}$, and the green line the realized excess market return. Vertical bars denote NBER-dated recessions.}
  \label{fig:forecasts}
\end{figure}

Figure~\ref{fig:forecasts} compares the two forecast series to realized returns. Forecasts based on $E^{PLS}$ are more volatile than those based on $E^{FC}$, while realized excess returns are most volatile. This difference is intuitive: PLS is a targeted method that loads more heavily on predictor components that covary with returns, whereas $E^{FC}$ averages across many individual predictive regressions, which mechanically smooths forecasts \parencite{Rapach2010}. The higher responsiveness of $E^{PLS}$ can be beneficial in environments where predictability is time varying.

\begin{figure}[h]
  \centering
  \includegraphics[width=0.8\linewidth]{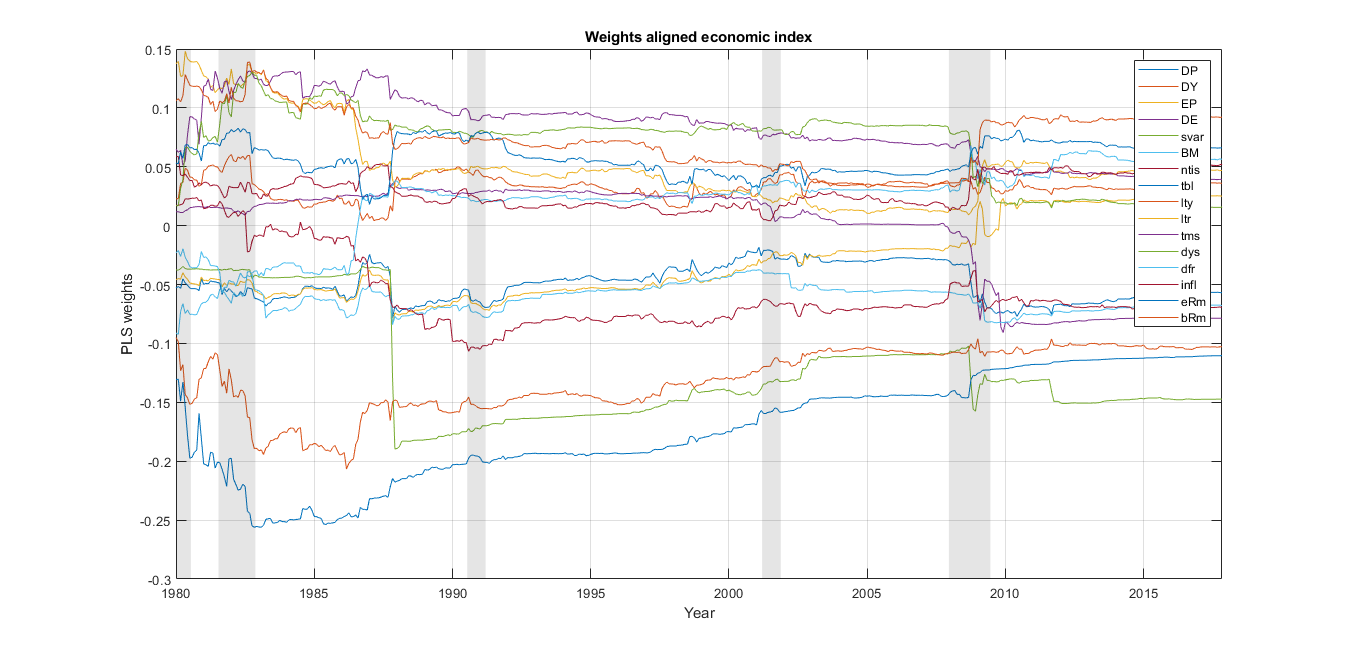}
  \caption{Weights of $E^{PLS}$ and $E^{FC}$ on fundamental variables, January 1980 to December 2017. The figure shows time-varying PLS weights for the 16 variables (DP, DY, EP, DE, SVAR, BM, NTIS, TBL, LTY, LTR, TMS, DFY, DFR, INFL, $eRm$, $eBm$). $E^{FC}$ assigns equal weights of 6.25\%. Vertical bars denote NBER-dated recessions.}
  \label{fig:weights}
\end{figure}

Figure~\ref{fig:weights} shows that $E^{PLS}$ places time-varying weight on the underlying predictors, with relatively large average weights on equity premium volatility, the Treasury bill rate, and the long-term yield. Weight variation is more pronounced around regime transitions, consistent with state-dependent and time-varying predictability \parencite{Devpura2018}. In contrast, $E^{FC}$ uses constant equal weights, which helps explain its smoother forecast path.

\section{Concluding remarks}

This paper studies return predictability using a real-time state-switching predictive regression and a new PLS-based aggregate predictor. The state-switching model uses the slope of the yield curve to define market regimes and generally improves the predictive performance of widely used predictors in sample and out of sample. The Aligned Economic Index $E^{PLS}$ aggregates information from the \textcite{WelchGoyal2008} predictor set (augmented with two additional series) using a targeted dimension-reduction approach. Empirically, $E^{PLS}$ delivers strong and robust forecasting performance, and its gains are especially pronounced when combined with the state-switching specification. Importantly, the resulting framework yields statistically significant and economically meaningful out-of-sample improvements across both expansionary and recessionary regimes, making it a practical alternative to one-state predictive regressions whose performance is often concentrated in recessions.

\printbibliography[heading=bibintoc, title=\ebibname]

\appendix
\addappheadtotoc

\section{Construction of the Aligned Economic Index}
\label{app:aei}

Assume the risk premium can be decomposed into a conditional expectation and an innovation:
\begin{equation}
R_{t+1} = \mathbb{E}_t[R_{t+1}] + \xi_{t+1}, \label{eq:app1}
\end{equation}
where $\mathbb{E}_t[R_{t+1}]$ is driven by an unobservable latent factor $F_t$. Suppose
\begin{equation}
\mathbb{E}_t[R_{t+1}] = \alpha_0 + \alpha_1 F_t. \label{eq:app2}
\end{equation}
Combining \eqref{eq:app1}--\eqref{eq:app2} yields
\begin{equation}
R_{t+1} = \alpha_0 + \alpha_1 F_t + \xi_{t+1}. \label{eq:app3}
\end{equation}
Next, assume predictors follow a factor structure:
\begin{equation}
x_{i,t} = \delta_{i,0} + \delta_{i,1} F_t + \delta_{i,2} Y_t + \eta_{i,t}, \quad i = 1,\dots,N, \label{eq:app4}
\end{equation}
where $x_{i,t}$ is predictor $i$, $Y_t$ is a common noise component, and $\eta_{i,t}$ is idiosyncratic noise. PCA extracts linear combinations of $x_{i,t}$ that explain total predictor variance, potentially loading on $Y_t$ and reducing forecasting usefulness. PLS instead targets the component most related to the forecast target.

In the first step, run $N$ time-series regressions of each lagged predictor on the realized risk premium (used as an instrument for the latent factor):
\begin{equation}
x_{i,t-1} = \phi_{i,0} + \phi_{i,1} R_t + \nu_{i,t}, \quad t=1,\dots,T. \label{eq:app5}
\end{equation}
Taking conditional expectations gives
\begin{equation}
\mathbb{E}_t[x_{i,t-1}] = \phi_{i,0} + \phi_{i,1}\mathbb{E}_t[R_t]. \label{eq:app6}
\end{equation}
Using $\mathbb{E}_t[R_t]=\alpha_0+\alpha_1 F_{t-1}$ implies
\begin{equation}
x_{i,t-1} = \phi_{i,0} + \phi_{i,1}(\alpha_0+\alpha_1 F_{t-1}) + \text{error}. \label{eq:app7}
\end{equation}
Thus $\phi_{i,1}$ captures how predictor $i$ loads on a rotation of the latent factor, while filtering out the unpredictable innovation component.

In the second step, for each $t$, run a cross-sectional regression of $\{x_{i,t}\}_{i=1}^N$ on the estimated first-stage loadings:
\begin{equation}
x_{i,t} = \phi_{0,t} + F_t \phi_i + \varphi_{i,t}, \quad i=1,\dots,N, \label{eq:app8}
\end{equation}
where $\phi_i$ denotes the estimated loading from the first stage. The fitted value yields the PLS factor estimate, which I denote by the Aligned Economic Index $E_t^{PLS}\equiv F_t^{PLS}$.

Finally, I use $E_t^{PLS}$ as the predictor in the state-switching regression:
\begin{equation}
\mathbb{E}_t[R_{t+1}] = (\beta_0 + \delta_0 S_t) + \beta_1 S_t E_t^{PLS} + \gamma_1(1-S_t)E_t^{PLS} + \varepsilon_{t+1}. \label{eq:app10}
\end{equation}
Following \textcite{KellyPruitt2015}, the second-stage factor estimate is a consistent estimator of the latent component relevant for forecasting. For the out-of-sample analysis, index weights and factor estimates are computed recursively using only information available up to the forecast-formation date.

\end{document}